\documentstyle[prl,aps,twocolumn]{revtex}

\catcode`\@=11

\def\maketitle2{\par 
\begingroup
\let\cite\@bylinecite
\def\thefootnote{\fnsymbol{footnote}}%
\twocolumn[\@maketitle2\vskip2pc]%
\thispagestyle{plain}\@thanks
\endgroup
\def\thefootnote{\arabic{footnote}}%
\setcounter{footnote}{0}%
\let\maketitle2\relax \let\@maketitle2\relax
\let\@thanks\relax \let\@authoraddress\relax \let\@title\relax
\let\@date\relax \let\thanks\relax \let\@abstract\relax
\let\@pacs\relax}

\def\abstract#1{\gdef\@abstract{{\par 
\bgroup
\ifdim\prevdepth=-1000pt \prevdepth0pt\fi
\hsize\columnwidth
\dimen0=-\prevdepth \advance\dimen0 by17.5pt \nointerlineskip
\small\vrule width 0pt height\dimen0 \relax}{~~}#1\egroup}}

\def\pacs#1{\gdef\@pacs{{\par 
\bgroup
\hsize\columnwidth \parindent0pt
\ifdim\prevdepth=-1000pt \prevdepth0pt\fi
\dimen0=-\prevdepth \advance\dimen0 by20pt\nointerlineskip
\egroup} PACS numbers:~#1}}

\def\@maketitle2{
\@preprint
\@title
\ifdim\prevdepth=-1000pt \prevdepth0pt\fi
\@authoraddress
\@date
\begin{list}{}{\leftmargin=0.10753\textwidth \rightmargin=\leftmargin
\itemsep=1pc\partopsep=-1pc}
\item\@abstract
\item\@pacs
\end{list}
}

\catcode`\@=12

\begin{document}

\draft
\title{Nonlinear Dynamics of a Bose Condensed Gas }
\author{F. Dalfovo, C. Minniti, S. Stringari}
\address{Dipartimento di Fisica, Universit\`a di Trento, and \\
Istituto Nazionale Fisica della Materia,
I-38050 Povo, Italy}
\author{L. Pitaevskii}
\address{Department of Physics, TECHNION, Haifa 32000, Israel,
and \\
Kapitza Institute for Physical Problems, ul. Kosygina 2, 117334 Moscow}

\date{December 18, 1996}

\abstract{We investigate the dynamic behavior of a Bose-condensed
gas of alkali atoms interacting with repulsive forces and
confined in a magnetic trap at zero temperature.
Using the Thomas-Fermi approximation, we rewrite the Gross-Pitaevskii
equation in the form of the hydrodynamic equations of superfluids.
We present solutions describing large amplitude oscillations of the
atomic cloud as well as the expansion of the gas after switching off
the trap. We compare our theoretical predictions with the recent
experimental data obtained at Jila and MIT.}


\maketitle2

\narrowtext

The dynamic behavior of Bose-condensed gases of alkali atoms in 
magnetic traps has been the object of recent experiments  
at Jila \cite{jila1,jila2,jila3} and MIT \cite{MIT1,MIT2,MIT3}. 
Absorption images of the  atomic cloud provide  
quantitative information on the dynamics  of the expansion 
after switching off  the trap as well as accurate data for 
the frequencies of the  collective excitations.  The experimental 
results reveal that the role played by the interatomic forces is 
important in these systems and can not be  ignored 
for a quantitative understanding of the data.
This opens new challenging tasks for theoretical investigation.

The natural theory to investigate the dynamic behavior of a nonuniform
Bose condensate, at $T=0$, 
is given by the time-dependent Gross-Pitaevskii (GP) 
equation \cite{GP} for the 
condensate wavefunction $\Psi({\bf r},t)$:
\begin{equation}
i\hbar\frac{\partial \Psi}{\partial t} =
\left( -\frac{\hbar^2\nabla^2}{2m} + V_{ext} 
+ g \mid \!\Psi\!\mid^2 \right) \Psi \, . 
\label{GP}
\end{equation}
The coupling constant $g$ is proportional to the  $s$-wave scattering 
length $a$ through $g=4\pi \hbar^2 a/m$. In the following we will 
discuss the case of repulsive interactions ($a>0$). The 
anisotropic trap is  represented by the confining potential $V_{ext}$, 
which is chosen in the form $V_{ext}({\bf r})= (m/2) \sum_i 
\omega_{0i}^2 r_i^2$, where $r_i \equiv x,y,z$. So far 
the experimental traps have cylindrical symmetry and hence 
are characterized by 
the radial frequency $\omega_\perp \equiv \omega_x=\omega_y$ and 
the  asymmetry parameter $\lambda=\omega_z/\omega_\perp$. 
The ground state configuration \cite{Baym,Dalfovo,Edwards1} as well 
as the properties of  small oscillations near equilibrium 
\cite{Edwards2,Singh,Stringari} have been the object of systematic 
investigation starting from Eq.~(\ref{GP}). A few calculations 
in the  nonlinear regime have been also carried out 
\cite{Holland1,Holland2,Castin,Kagan,Edwards3}. 

In the present paper we discuss several features of the
nonlinear behavior of the system, by solving Eq.~(\ref{GP})  in the large
$N$ limit, where $N=\int\!d{\bf r} |\Psi({\bf r},t)|^2$ is the number 
of atoms.  In this limit, it is
possible to derive almost analytic results of the GP equation,
thereby simplifying the numerical analysis and allowing for  a 
systematic investigation of important phenomena. 
These include, the explicit time evolution of the condensate (shape 
of profiles, aspect ratio etc.) during the expansion and the dependence
of the collective frequencies on the amplitude of the oscillation.

The effective strength of the interatomic forces in the 
GP equation is fixed by the adimensional parameter $Na/a_{0i}$, 
where $a_{0i}= \sqrt{\hbar/(m\omega_{0i})}$ is the harmonic oscillator 
length. When this parameter is much larger than $1$, the repulsion 
makes the system much wider than the noninteracting configuration, 
yielding a rather smooth profile.  In such conditions, the equilibrium 
results from a balance between the  external potential and the 
repulsive interaction, the kinetic energy playing a minor role.
For large values of $Na/a_{0i}$, one can then neglect the kinetic 
energy term in (\ref{GP}). This yields the  Thomas-Fermi (TF) 
approximation for the ground state:
\begin{equation}
\rho_0^{TF}({\bf r}) = | \Psi_0^{TF} ({\bf r})|^2 
= g^{-1} [ \mu - V_{ext}({\bf r}) ]
\label{tfgs}
\end{equation}
when $\mu > V_{ext}({\bf r})$ and $\rho_0({\bf r}) =0$ elsewhere. The
chemical potential $\mu$ is fixed by the normalization of the density
to the number of particle $N$:
\begin{equation}
\mu = {1\over 2} \left[ {15 \over 4\pi} g m^{3/2} \omega_{0x}
\omega_{0y } \omega_{0z } N
\right]^{2/5} \; .
\label{mu}
\end{equation}
The Thomas-Fermi approximation (\ref{tfgs}) works in an excellent way 
for the configurations realized at MIT, where $N$ is of the order 
of 1 million atoms and more. Conversely, in the Jila experiments 
of Refs.~\cite{jila1,jila2,jila3}
the number of atoms is smaller ($10^3$-$10^4$) and the TF 
approximation provides only a semi-quantitative description.

Neglecting the kinetic energy pressure term in the time dependent 
GP equation corresponds to taking into account the  effect of the 
kinetic energy operator in (\ref{GP}) only on the phase of the order 
parameter $\Psi$. This permits to rewrite (\ref{GP}) in the useful 
hydrodynamic form \cite{Stringari}:
\begin{eqnarray}
\frac{\partial}{\partial t} \rho &+& {\bf \nabla} \cdot ({\bf v}\rho) 
= 0 
\label{continuity} \\
m \frac{\partial}{\partial t} {\bf v} &+&
{\bf \nabla} \left( V_{ext} + g\rho 
+ { mv^2 \over 2 } \right) = 0 \; ,
\label{Euler} 
\end{eqnarray}
where density and velocity are defined by 
$\rho=|\Psi|^2$ and   
${\bf v}= (\Psi^* {\bf \nabla}\Psi -\Psi{\bf \nabla}\Psi^*)
\hbar/(2mi\rho)$.  Equation (\ref{continuity}) is the usual equation 
of continuity, while (\ref{Euler}) establishes 
the irrotational nature of the  superfluid velocity.
It is immediate to verify that the equilibrium configuration
given by Eqs.~(\ref{continuity}-\ref{Euler}) (${\bf v} =0$ and
$\partial \rho / \partial t =0$) coincides with the TF result
(\ref{tfgs}). 

The hydrodynamic equations (\ref{continuity}-\ref{Euler}) have been
recently shown to provide the correct frequencies of the normal modes of 
the condensate in the large $N$ limit \cite {Stringari}. 
With respect to the full solution of the GP equation, which 
includes the effect of the kinetic energy
pressure, the approach based on (\ref{continuity}-\ref{Euler}) has the
major advantage of providing an algebraic expression  for the
dispersion relation of the elementary excitations. The resulting predictions
quite well compare with both the Jila \cite{jila2,jila3} 
and MIT \cite{MIT3} experiments. 

For nonlinear time dependent motions, which are the object of the present 
work, almost analytic solutions can be found starting from  equations 
(\ref{continuity}-\ref{Euler}). In fact they admit {\sl exact} solutions 
of the form
\begin{eqnarray}
\rho({\bf r},t) &=& a_x(t)  x^2 + a_y(t) y^2 + a_z(t) z^2 + a_0(t)
\label{scalingrho} \\
{\bf v}  &=& {1 \over 2} {\bf \nabla} [\alpha_x(t) x^2 + \alpha_y(t) y^2 
+ \alpha_z(t) z^2] \; .
\label{scalingv}
\end{eqnarray}
Equation (\ref{scalingrho}) is restricted to the region where 
$\rho\ge 0$ and the coefficient $a_0$ is fixed by the normalization of 
the density: $a_0=-(15N/8\pi)^{2/5}(a_xa_ya_z)^{1/5}$. 
The time dependent coefficients $a_i$ and $\alpha_i$ obey  
the following coupled differential equations: 
\begin{eqnarray}
\dot{a}_i &+& 2a_i\alpha_i+a_i\sum_j\alpha_j =0 
\label{ai} \\
\dot{\alpha}_i &+& \alpha_i^2+\omega_{0i}^2+(2g/m) a_i = 0 \; ,
\label{alphai}
\end{eqnarray}
with $i,j=x,y,z$. 
One can use the three equations
(\ref{ai}) to express $\alpha_i$ in terms of $\dot{a}_i/a_i$; the
solution is greatly simplified by introducing the new adimensional
variables $b_i$ defined by $a_i=-m\omega_{0i}^2 (2g b_x b_y b_z
b_i^2)^{-1}$. With this choice, Eqs.~(\ref{ai}) reduce to 
$\alpha_i=\dot{b}_i/b_i$ and Eqs.~(\ref{alphai}) become
\begin{equation}
\ddot{b}_i  + \omega_{0i}^2 b_i - \omega_{0i}^2 /(b_i b_x b_y b_z) 
= 0 \; .
\label{ddotb}
\end{equation}
The second and third terms of (\ref{ddotb}) give 
the effect of the external trap and of  the interatomic forces, 
respectively. It is worth noticing that, using the new
variables $b_i$, the equations of motion do 
not depend on the value of the coupling constant $g$. This is 
a typical feature characterizing the large $N$ behavior of the 
GP equation. The mean square radii and velocities of the atomic 
cloud can be easily expressed in terms of $b_i$:
\begin{eqnarray}
\langle r^2_i \rangle & \equiv & {1\over N} \int \! d{\bf r} \ 
\rho({\bf r},t) \  r^2_i \  =  
\left( {2 \mu \over 7 m \omega_{0i}^2 }\right)  b_i^2  \\
\langle v^2_i \rangle & \equiv & {1\over N} \int \! d{\bf r} \
\rho({\bf r},t) \  v^2_i \  =
\left( {2 \mu \over 7 m \omega_{0i}^2 }\right)  \dot{b}_i^2  \; ,
\label{rms} 
\end{eqnarray}
where $\mu$ is given by (\ref{mu}). 

The solutions (\ref{scalingrho}-\ref{scalingv}) are 
well suited to describe both the oscillations around the ground 
state \cite{note1}  and the problem of the expansion of the gas
after switching off the confining potential.
Equations (\ref{ddotb}) have been already derived by 
other authors \cite{Castin,Kagan}, using the formalism of scaling 
transformations,  and applied to study several dynamic phenomena.
In the present approach the same equations emerge 
as an exact solution of the hydrodynamic equations of
superfluids (\ref{continuity}-\ref{Euler}) and 
will be used to investigate nonlinear oscillations and the expansion 
in both the Jila and MIT traps. 

In order to apply the above formalism to study the expansion of the 
cloud, let us suppose that at $t=0$ the system is in its equilibrium 
configuration. Comparing Eq.~(\ref{scalingrho}) with the 
TF ground state density (\ref{tfgs}) and substituting $a_i$ in terms
of $b_i$, one finds $b_i=1$. One has also $\dot{b}_i=0$ at equilibrium. 
The external potential is then suddenly switched off and the
system starts expanding. This corresponds to solving the equations 
(\ref{ddotb}) with the second term set equal to zero:
\begin{equation}
\ddot{b}_i  - \omega_{0i}^2 /(b_i b_x b_y b_z)
= 0 \; .
\label{expansion}
\end{equation}

For an initially spherical configuration the expansion will proceed
isotropically. In the presence of anisotropy, the expansion 
(and consequently the asymptotic velocities) will be instead faster
in the direction where the repulsive forces (proportional to the gradient
of the density) are stronger. For an axially deformed trap with $\lambda
= \omega_z/\omega_\perp  \ge 1$ this will occur in the axial direction, 
while if $\lambda \le 1$ it will occur in the radial direction.
The ratio
\begin{equation}
R_r(t) \equiv \sqrt{ \langle z^2 \rangle / \langle x^2 \rangle  } 
=    \lambda^{-1} b_z(t)/b_x(t)   
\end{equation}
of the radii in the two  different directions is called the aspect ratio 
in co-ordinate space. One can also define the aspect ratio of velocities 
\begin{equation}
R_v(t) \equiv \sqrt{  \langle v_z^2 \rangle / \langle v_x^2 \rangle }
=  \lambda^{-1} \dot{b}_z(t) / \dot{b}_x(t)  \; , 
\end{equation}
whose deviations from unity reflect the anysotropy of the velocity 
distribution. This anysotropy represents a crucial feature of Bose 
condensates.  Asymptotically ($t \to \infty$) the two aspect ratios 
$R_r$ and $R_v$  converge to the same value $R$. 
However  for finite values of the expansion time
they can behave quite differently. In fact, the velocities in both
radial and axial directions reach their asymptotic values very rapidly,
since the time scale for acceleration is very short; vice versa 
the radii approach their asymptotic behavior, 
$\propto v_i t$,  much more slowly.
This is an important feature to take into account in the analysis 
of experimental data.

In Figs.~\ref{fig:jila} and \ref{fig:mit} we show the results of the 
two aspect ratios $R_r$ and $R_v$ obtained by solving numerically 
Eqs.~(\ref{expansion})  for two different sets of frequencies 
$\omega_{0i}$, corresponding to the Jila \cite{jila2,jila3} 
and MIT \cite{MIT2} traps, respectively. The  different behavior 
of the two aspect ratios $R_r$ and $R_v$ is evident in both cases.  
In the same figures we have also shown the predictions of the
noninteracting harmonic oscillator model, which gives
$\langle v_i^2 \rangle = \hbar \omega_{0i}/(2m)$ and 
$\langle r^2_i \rangle =  (1/2) a_{0i}^2 + \langle v_i^2 \rangle t^2$. 
The figures point out very clearly the role of two-body interactions
which modify both the timescale of the expansion process and the
asymptotic value of the aspect ratio. The comparison with the experimental 
data for $R_r$ should be however taken with care. In the case of the Jila 
experiments, $N$ is of the order of $10^3\div 10^4$,  
so that the TF approximation
is expected to be rather crude; moreover, the  points in 
Fig.~\ref{fig:jila} are taken from a gaussian fit to the spatial
distribution of the atoms and they can be significantly corrected by 
using different fitting functions \cite{Holland2,Jin}. In the case of the 
MIT experiments, the points in Fig.~\ref{fig:mit} correspond to our 
estimate of the aspect ratio, extracted from the time-of-flight 
images in Fig.~1 of Ref.~\cite{MIT2}. 

In Fig.~\ref{fig:lambda} we report the asymptotic aspect ratio $R$ as a 
function of $\lambda$. This curve 
has been recently calculated also in Ref.~\cite{Kagan}. In the 
limit $\lambda \to 0$  the aspect ratio approaches the value 
$(\pi/2) \lambda$ \cite{Castin}. 
For comparison we also show the predictions of the noninteracting 
harmonic oscillator model, given by $R_{HO} = \sqrt{\lambda}$. 

Another important quantity to discuss is the release energy 
$E_{rel}$, defined as the energy per particle of the system after the 
switch-off of the trap.  This energy is given by the 
sum of the kinetic and interaction  energy  of the atoms and it
is conserved during the expansion, being finally converted 
entirely into the kinetic energy of the expanding cloud. 
In the present formalism the release energy is a first integral of 
equation (\ref{expansion}) and can be written as
\begin{equation}
E_{rel} = {2 \mu \over 7} \left[ {1 \over b_xb_yb_z} +
{1\over2} \sum_i { \dot{b}_i^2 \over \omega_{0i}^2 }\right] 
= {2 \mu \over 7} \; ,
\label{ereltf}
\end{equation}
where $\mu$ is given by (\ref{mu}).  At the beginning of the 
expansion $E_{rel}$ coincides with the 2-body interaction energy of the 
TF ground state.  The comparison between Eq.~(\ref{ereltf})  and the 
numerical results for $E_{kin}+E_{int}$, calculated with the exact 
ground state solution of the Gross-Pitaevskii equation, provides a 
test of the validity of the TF approximation. For $N \simeq 10^3$
the agreement with the exact result is only semiquantitative, becoming 
better and better as $N$ increases \cite{Dalfovo}. After long expansion 
time  the release energy can be related to the  value of the square 
radii of the system, through the equation 
\begin{equation}
E_{rel} = {m \over 2} \langle v^2 \rangle_{t \to \infty} =
{m \over 2 t^2} \langle r^2 \rangle_{t \to \infty} \; . 
\label{erel}
\end{equation}
The release energy has been measured both at 
Jila \cite{Holland2} and  MIT \cite{MIT2}. 
It is worth noticing that in the MIT trap, where 
the initial configuration is a strongly anisotropic ellipsoid with 
the major axis along $z$, the release energy is almost entirely 
converted into kinetic energy of the radial motion, the 
velocity along $z$ being much smaller. This can be 
tested by solving equation (\ref{expansion}) with the  
parameters appropriate for the trap of  Ref.~\cite{MIT2}.
One finds that, after an 
expansion of $40$ ms, the 2-body interaction energy is a 
factor $10^{-4}$  smaller than the initial value and the ratio 
between the axial and radial kinetic energies is approximately 
$4 \times 10^{-3}$.  

The same formalism can be used to investigate the oscillations of the
trapped gas. One can easily check that, in the limit of small
deformations, the solutions of (\ref{ddotb}) yield the 
dispersion relation discussed in Ref.~\cite{Stringari}. For an 
axially deformed trap the normal modes are classified in terms of
the third component  $m$ of the angular momentum. We will discuss 
here the $m=0$ and $m=2$ modes which are accounted for by the 
parametrization (\ref{scalingrho}-\ref{scalingv}). The 
$m=2$ mode, in the linear limit, has the frequency $\omega=\sqrt{2}
\omega_\perp$, while the low-lying  $m=0$ mode, resulting from the coupling 
between {\it monopole} and {\it quadrupole} oscillations, has the 
frequency  \cite{Stringari} $\omega^2 = \eta \omega_{\perp}^2$, with 
$\eta= (4 + 3 \lambda^2 - \sqrt{9\lambda^4-16\lambda^2+16} )/2$.
The frequency of the collective modes is expected to change when the 
amplitude of the oscillations becomes large, due to nonlinear effects. 
In order to calculate such deviations, we solve (\ref{ddotb}) using,
as initial conditions, the ground state values $b_i(0)=1$, but with
$\dot{b}(0) \ne 0$. We choose the values of the velocities 
$\dot{b}_i(0)$ in order to to excite, in the linear limit, the two separate 
$m=2$ and  $m=0$ modes. For the $m=2$ mode this implies $\dot{b}_x=-
\dot{b}_y=\epsilon$ and $\dot{b}_z=0$, where the parameter $\epsilon$ 
fixes the amplitude of the oscillations. For the $m=0$ mode, one has 
$\dot{b}_x=\dot{b}_y=\epsilon$ and $\dot{b}_z=\epsilon(\eta-4)$.
In this case, the system oscillates also along $z$, the axial and
radial oscillations having relative amplitude $\eta-4$. 
It is worth noticing that the occurrence of a simultaneous 
oscillation in both the radial and axial widths is a typical
effect of the interaction between the atoms. In fact, in the 
absence of 2-body forces (noninteracting harmonic oscillator), the 
motion in the two directions would be exactly decoupled. The experiments
of Ref.~\cite{jila2} reveal not only a good agreement with the 
predicted frequencies of the two modes but also a clear evidence 
for the coupling between the axial and radial 
oscillations (see Fig.~2 of Ref.~\cite{jila2}). These results are 
crucial signatures of the important role played by the interaction.

By increasing the initial values $\dot{b}_i$, it is possible to 
explore the nonlinear regime. A major problem for the comparison with 
the experimental data is that the oscillations are measured 
by imaging the atomic cloud after switching off the trap and leaving 
the atoms to expand for a few ms. During the expansion the 
relative amplitude of the axial and radial motions can be significantly
modified. This is especially true for the MIT trap, due to the 
strong asymmetry of the starting cigar-shaped configuration which makes
the expansion in the radial and axial directions quite different. 
This nontrivial evolution of the oscillating cloud 
is not surprising if one thinks that, after an expansion time
of a few ms, the size of the system can increase by more than a factor 
ten. In order to make a significant comparison with the experiments 
it is then important to simulate both the oscillations in the trap and 
the subsequent expansion. 

In Fig.~\ref{fig:frequencies} we show our predictions for the frequencies
of the $m=0$ and $m=2$ modes, in the case of the Jila trap ($\lambda=
\sqrt{8}$), as a function of the relative amplitude,
defined as $(1/2)[\langle x^2 \rangle_{max}^{1/2} -  \langle x^2 
\rangle_{min}^{1/2}]/ \langle x^2 \rangle_{ave}^{1/2}$. The latter
is calculated after expanding the trap for $7$ ms, as in the experiments
of Refs.~\cite{jila2,jila3}. We find that the frequency of the 
$m=0$ mode does not exhibit any significant dependence on the relative
amplitude, while the $m=2$ frequency increases. This agrees with the 
experimental findings \cite{jila2,jila3}, though the measured 
frequency shift of the $m=2$ mode is about a factor two larger 
than our prediction. 

As already pointed out, the relative amplitude of the 
oscillations in the radial and axial
directions behave differently during the expansion. For example, 
after exciting the $m=2$ mode in the trap, with a relative amplitude
of $5$\%\ and $4$\%\ in the radial and axial directions respectively,
the corresponding oscillations of the expanded cloud, after $7$ ms,  
have a relative amplitude of $9$\%\ and $4$\%. 
The aspect ratio $R_r$ after $7$ ms is found to be $1.65$. These 
predictions rather well 
agree with the experiments (see Fig.~2 of Ref.~\cite{jila2}), where 
the aspect ratio is found to be $1.75$ and the radial and axial 
relative amplitudes are $10$\%\ and $4$\%, respectively. 

We have repeated the same calculations for the cigar-shaped trap of 
the MIT experiments ($\lambda=0.077$ \cite{MIT3}). In this case 
we excite the low-lying $m=0$ mode in the trap and then we calculate 
the frequency and the  amplitude of the oscillations after expanding 
the gas for $40$ ms. We do not observe any frequency shift of the 
low-lying $m=0$ mode as a function of the amplitude, in agreement with 
the experiments. One should note that the nonlinearity of the oscillations 
are amplified during the expansion much more than in the Jila trap, since
the system is very anisotropic. As a consequence, the large axial amplitude
observed after an expansion of $40$ ms corresponds to a rather small
amplitude at $t=0$ and this explain in part the absence of shift in 
the frequency.  As an example, let us consider the aspect ratio plotted 
in Fig.~2 of Ref.~\cite{MIT3}, which oscillates between about $0.28$ 
and $0.38$. One can easily reproduce the same 
oscillations by solving (\ref{expansion}) starting from a configuration 
which oscillates as a low-lying $m=0$ mode in the trap (see also 
Ref.~\cite{Castin}). To get the measured aspect ratio one has to start
with an oscillation having  relative  amplitude less 
than $1$\%\ in the radial direction and about $3.3$\% 
in the axial one.  During the expansion the relative amplitude of the 
radial motion remains practically unchanged while the one of the axial 
motion increases up to $15$ \%. These large fluctuations of the axial 
width of the system are compatible with the conservation of energy, 
since almost all the energy is carried by the motion of the atoms
in the radial expansion, which is much faster than the axial one.  

It is also worth noticing that, when $\lambda \ll 1$, the frequency of 
the high-lying $m=0$ mode is exactly $2\omega_\perp$ even in the 
nonlinear regime. In this limit, this mode corresponds to a 
two-dimensional motion and the absence of a shift in its frequency  
reflects the occurrence of a hidden  symmetry of a 2D Bose gas in a 
harmonic trap \cite{Pit}. 

In conclusion we have shown that the hydrodynamic equations of superfluids
provide a useful description of several nonlinear effects, 
associated with the dynamic behavior of a trapped Bose gas at zero
temperature. A reasonable agreement with the first available 
experimental data is found, though for a more quantitative comparison,
when the number of atoms is relatively small, the complete solution
of the Gross-Pitaevskii equations is expected to be relevant. Our 
analysis points out the crucial role played by the interatomic forces 
in the dynamics of the expansion as well as in the behavior of the collective 
excitations. A natural extension of this work should include thermal
effects and, in particular, the interaction between the consensed and
thermal components of these systems.

\begin{figure}
\caption{Aspect ratio for the Jila trap ($\omega_\perp=2\pi 132$ Hz and
$\protect\lambda=\protect\sqrt{8}$) as a function of the expansion time. 
Upper and lower solid lines correspond to $R_v$ and $R_r$, respectively.
Dashed lines are the same quantity for the noninteracting particles. 
Points are experimental values of $R_r$ taken from \protect\cite{Jin}. 
}
\label{fig:jila}
\end{figure}

\begin{figure}
\caption{Aspect ratio for the MIT trap ($\omega_\perp=2\pi
320$ Hz and $\lambda= 0.056$)  as a function of the expansion time.
Lower and upper solid lines correspond to $R_v$ and $R_r$, respectively.
Dashed lines are the same quantity for the noninteracting particles.
Points are experimental values of $R_r$ extracted from 
Ref.~\protect\cite{MIT2} (see text). 
}
\label{fig:mit}
\end{figure}

\begin{figure}
\caption{Asymptotic aspect ratio {\it vs.} $\lambda=\omega_z/\omega_\perp$. 
Solid lines: from (\protect \ref{expansion}); dashed line: noninteracting gas.
}
\label{fig:lambda}
\end{figure}

\begin{figure}
\caption{ Frequencies of the collective modes as a function of the
relative radial amplitude, for the $m=0$ (triangles) and $m=2$ (circles)
modes in the Jila trap. Solid lines: from (\protect \ref{ddotb}); 
points: from Refs.~\protect \cite{jila2,jila3}. }
\label{fig:frequencies}
\end{figure}


\begin{references}

\bibitem{jila1} M.\ H.\ Anderson, J.\ R.\ Ensher,
M.\ R.\ Matthews, C.\ E.\ Wieman, and E.\ A.\ Cornell,
Science {\bf 269}, 198 (1995).

\bibitem{jila2} D.\ S.\ Jin, J.\ R.\ Ensher, M.\ R.\ Matthews,
C.\ E.\ Wiemann, and E.\ A.\ Cornell,
Phys. Rev. Lett. {\bf 77}, 420 (1996).

\bibitem{jila3} D.\ S.\ Jin,  M.\ R.\ Matthews, J.\ R.\ Ensher,
C.\ E.\ Wiemann, and E.\ A.\ Cornell, preprint 1996

\bibitem{MIT1} K.\ B.\ Davis, M.-O.\ Mewes, M.\ R.\ Andrews,
N.\ J.\ van Druten, D.\ S.\ Durfee, D.\ M.\ Kurn, and W.\ Ketterle,
Phys. Rev. Lett. {\bf 75}, 3969 (1995).

\bibitem{MIT2} M.-O.\ Mewes, M.\ R.\ Andrews,
N.\ J.\ van Druten, D.\ M.\ Kurn, D.\ S.\ Durfee,
and W.\ Ketterle, Phys. Rev. Lett. {\bf 77}, 416 (1996)

\bibitem{MIT3} M.-O.\ Mewes, M.\ R.\ Andrews,
N.\ J.\ van Druten, D.\ M.\ Kurn, D.\ S.\ Durfee, C.\ G.\ Townsend,
and W.\ Ketterle, Phys. Rev. Lett. {\bf 77}, 988 (1996)

\bibitem{GP}  L.P.~Pitaevskii, Zh. Eksp. Teor. Fiz. {\bf 40}, 646 (1961)
[Sov. Phys. JETP {\bf 13}, 451 (1961)]; E.P.~Gross, Nuovo Cimento
{\bf 20}, 454 (1961); E.P.~Gross, J. Math. Phys. {\bf 4}, 195 (1963).

\bibitem{Baym} G.\ Baym and C.\ Pethick,
\prl {\bf 76}, 6 (1996).

\bibitem{Dalfovo} F. Dalfovo and S. Stringari, Phys. Rev. A {\bf 53},
4377 (1996).

\bibitem{Edwards1} Mark Edwards, R.\ J.\ Dodd,
C.\ W.\ Clark, P.\ A.\ Ruprecht, and K.\ Burnett,
Phys. Rev. A {\bf 53}, R1950 (1996). 

\bibitem{Edwards2} Mark Edwards, P.\ A.\ Ruprecht,
K.\ Burnett, R.\ J.\ Dodd, and C.\ W.\ Clark,
\prl {\bf 77}, 1671 (1996).

\bibitem{Singh} K.G. Singh and D.S. Rokhsar, Phys. Rev. Lett.
{\bf 77}, 1667 (1996) 

\bibitem{Stringari} S. Stringari, Phys. Rev. Lett. {\bf 77}, 2360 (1996)

\bibitem{Holland1} M.\ J.\ Holland and J.\ Cooper,
\pra {\bf 53}, R1954 (1996).

\bibitem{Holland2} M.J. Holland, D. Jin, M.L. Chiofalo, and 
J. Cooper, preprint 1996

\bibitem{Castin} Y. Castin and R. Dum, preprint 1996

\bibitem{Kagan} Yu. Kagan, E.L. Surkov and G.V. Shlyapnikov, Phys.
rev. A {\bf 54}, R1753 (1996); Yu. Kagan, E.L. Surkov and G.V. Shlyapnikov,
preprint

\bibitem{Edwards3} P.A. Ruprecht, M. Edwards, K. Burnett, and C.W. Clark,
Phys. Rev. A {\bf 54}, 4178 (1996)

\bibitem{note1} Another exact solution of 
Eqs.~(\protect \ref{continuity}-\ref{Euler}) is obtained including terms 
linear in $x$ (or $y, z$) in the density and in the velocity potential.
This solution corresponds to the motion of the center of mass of the cloud.
Other solutions can be found including terms of the form $xy$ ($xz$, $yz$).

\bibitem{Jin} D.\ S.\ Jin and E. A. Cornell, private communication

\bibitem{Pit} L.P. Pitaevskii, Phys. Lett. {\bf A 221}, 14 (1996); L.P. 
Pitaevskii and A. Rosch, preprint cond-mat/9608135


\end{references}
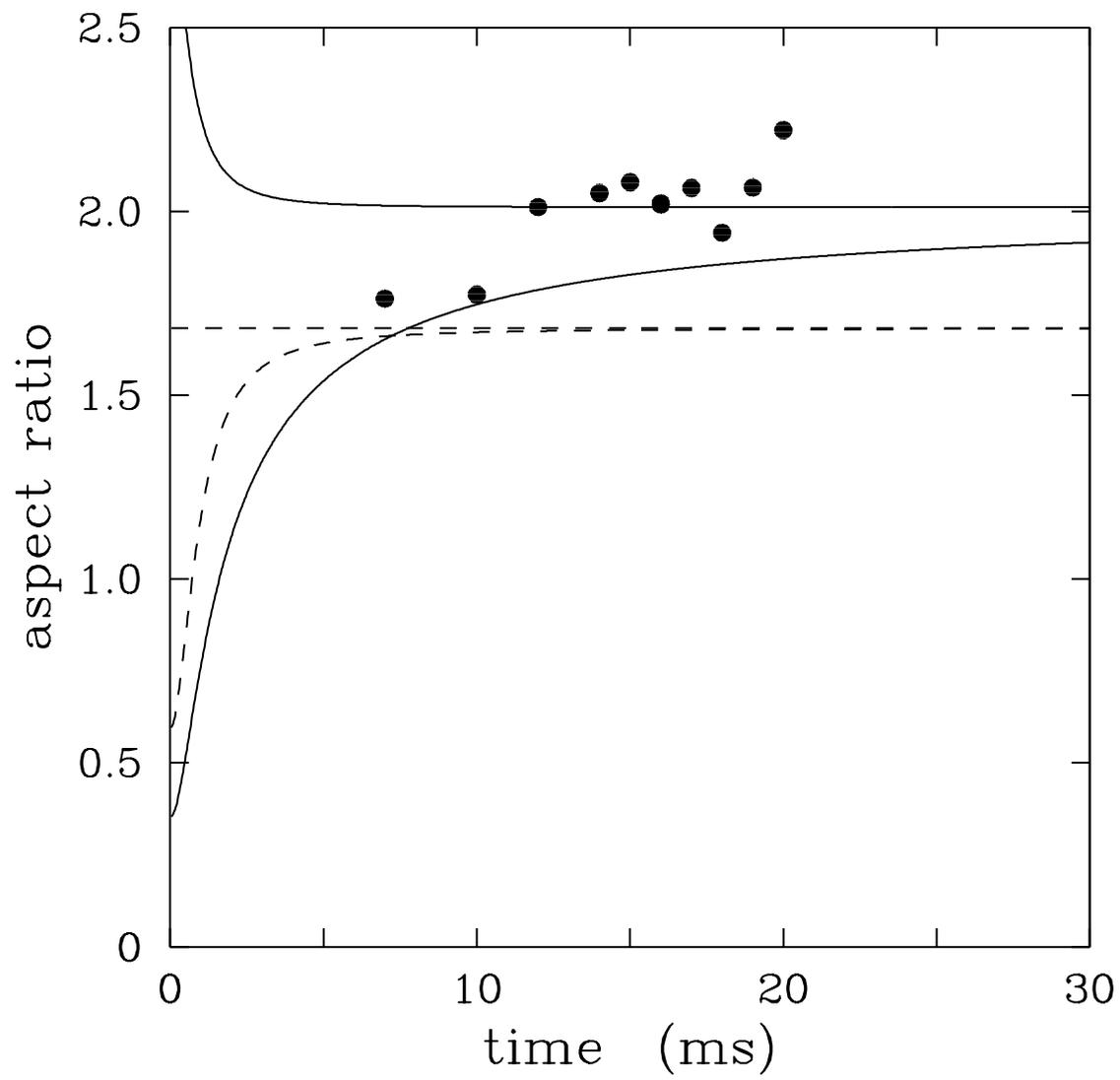
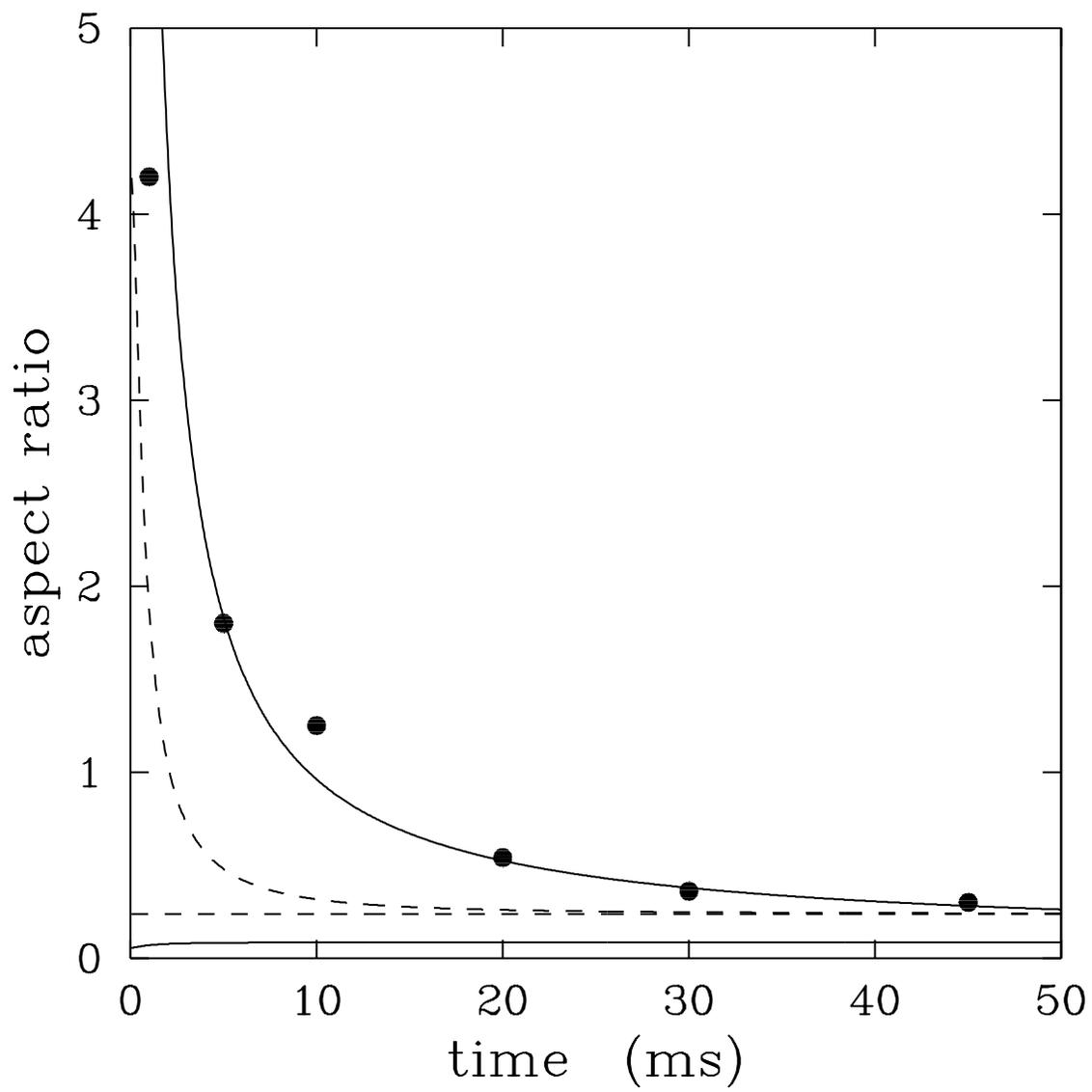
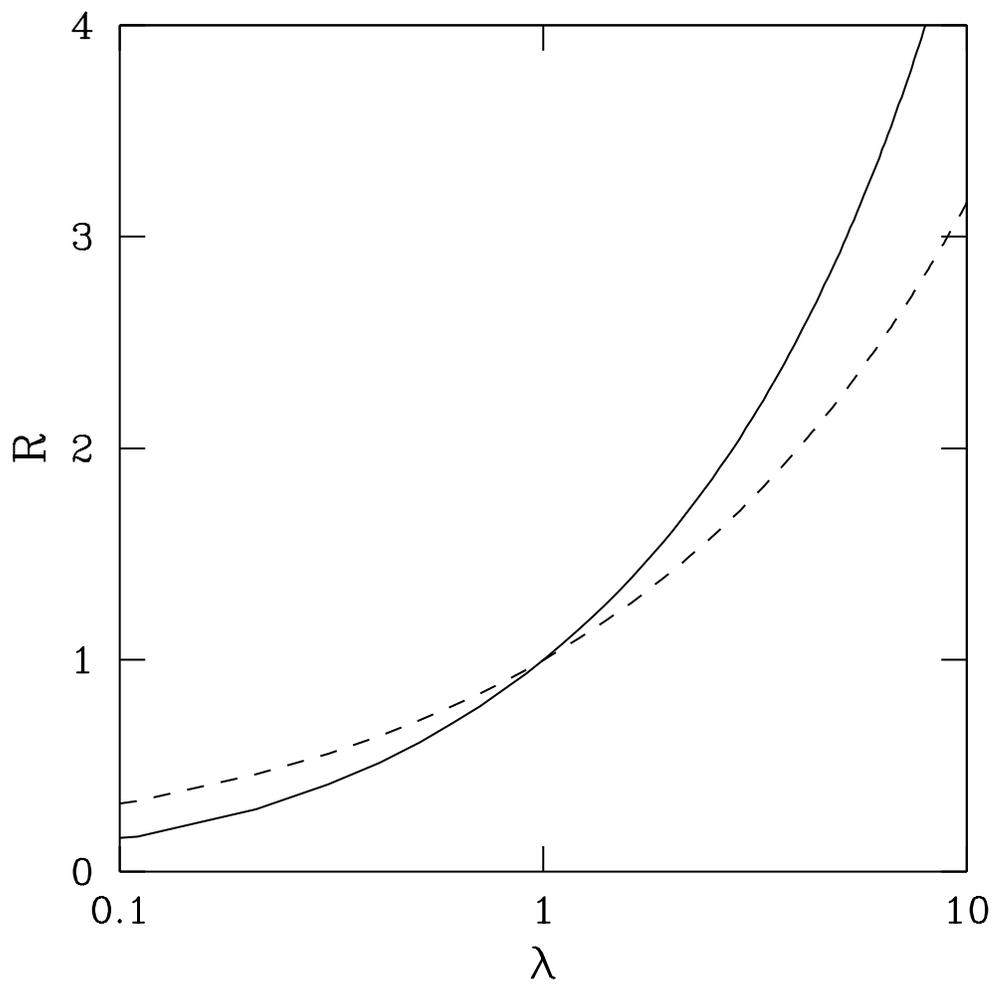
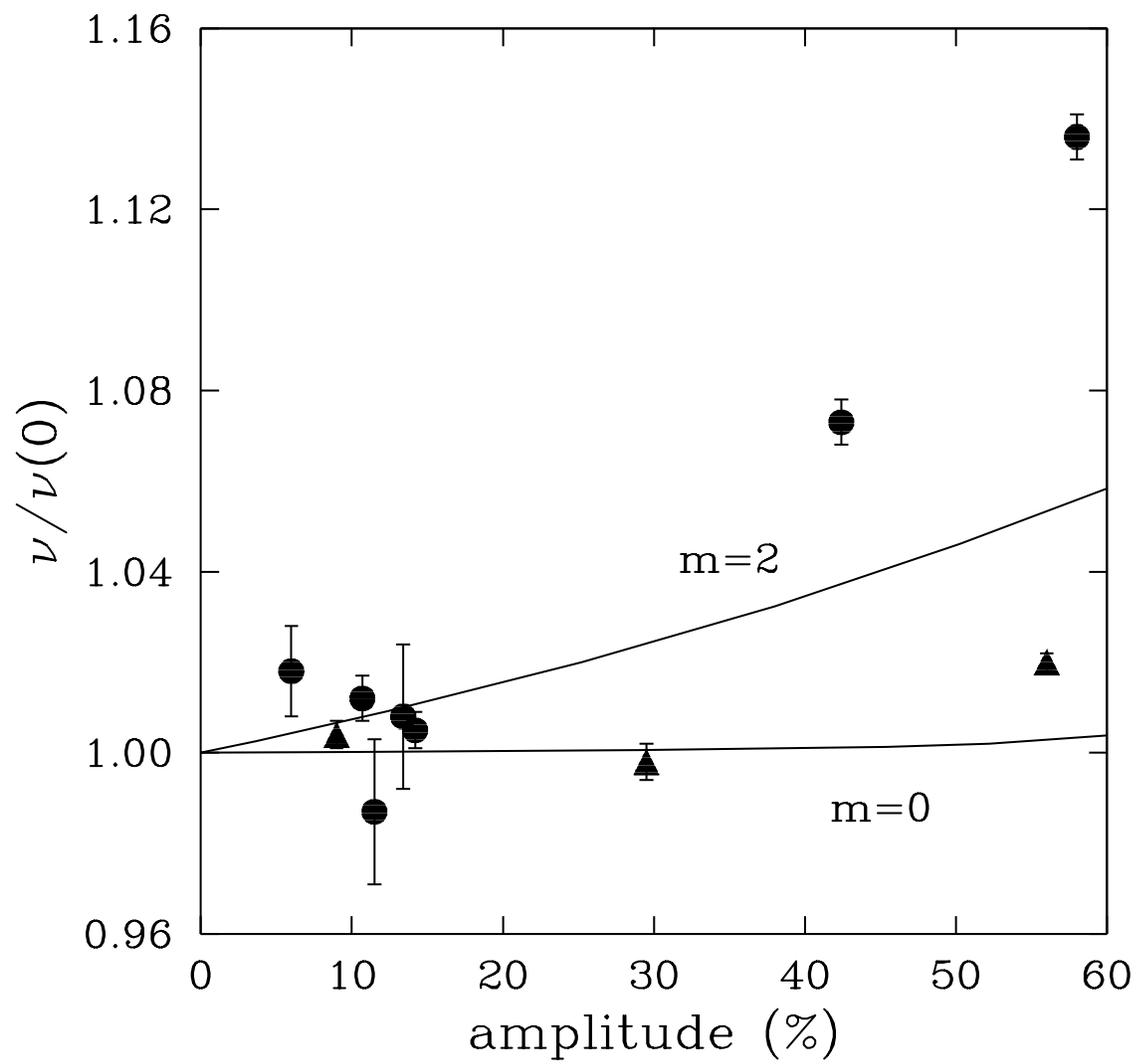
\end{document}